\documentclass[fleqn,12pt,twoside]{article}
\usepackage{espcrc1}

\usepackage{graphicx}
\usepackage[figuresright]{rotating}

\newcommand{\AmS}{{\protect\the\textfont2
  A\kern-.1667em\lower.5ex\hbox{M}\kern-.125emS}}

\newcommand{\zr}[1]{\mbox{\hspace*{#1em}}}

\newcommand{\ZZ}{\mbox{\sf Z\zr{-0.45}Z}}

\newcommand{\Fig}[1]{Fig.~\protect\ref{#1}}

\title{Penta-quark in Anisotropic Lattice QCD\thanks{
The lattice QCD numerical calculation has been done on NEC SX-5
at Osaka University.}}

\author{
N.~Ishii\address[titech]{
Department of Physics, Tokyo Institute of Technology,
Tokyo 152-8551, Japan
},
T.~Doi\address[RBRC]{
RIKEN BNL Research Center, Brookhaven National Laboratory,
NY 11973, USA
},
H.~Iida\addressmark[titech]{},
M.~Oka\addressmark[titech]{},
F.~Okiharu\address[nichidai]{
Department of Physics, Nihon University,
Chiyoda, Tokyo 101-8308, Japan
}
and
H.~Suganuma\addressmark[titech]{}
}
       
\begin{document}

\maketitle

\begin{abstract}
Penta-quark (5Q) baryons are studied using anisotropic lattice QCD for
high-precision measurement of temporal correlators.
A non-$NK$-type interpolating field is employed to study the 5Q states
with  $J^{P}=1/2^{\pm}$  and  $I=0$.   In $J^{P}=1/2^+$  channel,  the
lowest-lying state is  found at $m_{\rm 5Q}\simeq 2.25$  GeV, which is
too   massive  to   be   identified  as   the  $\Theta^+(1540)$.    In
$J^{P}=1/2^-$  channel, the  lowest-lying  state is  found at  $m_{\rm
5Q}\simeq 1.75$ GeV.
To distinguish  a compact 5Q  resonance state from an  $NK$ scattering
state, a new method with  ``{\em hybrid boundary condition (HBC)}'' is
proposed.  As a result of the  HBC analysis, the observed state in the
negative-parity channel turns out to be an $NK$ scattering state.
\end{abstract}

\section{Introduction}
The discovery of a  narrow resonance $\Theta^+(1540)$ has an important
impact on the studies  of exotic hadrons \cite{nakano}. This resonance
is peaked at $1.54\pm 0.01$ GeV  with a decay width narrower than $25$
MeV. It  is confirmed to have  baryon number $B=1$,  charge $Q=+1$ and
strangeness  $S=1$.   The  quantum  number  itself  implies  that  the
$\Theta^+$ has to contain at least one $\bar{s}$.  Hence, its simplest
configuration is $uudd\bar{s}$,  i.e., a manifestly exotic penta-quark
(5Q).   The  experimental discovery  was  motivated  by a  theoretical
prediction \cite{diakonov}.

Since its discovery, there have been an enormous number of theoretical
contributions \cite{oka,zhu}.
The determination  of its parity is an  important topic.  Experimental
determination  of  parity of  the  $\Theta^+$  is  difficult, and  the
opinions are divided into two pieces in the theoretical side.
There   are   several   lattice    QCD   studies   of   5Q   available
\cite{csikor,sasaki,chiu,mathur,ishii,takahashi}, which, however, have
not yet reached a consensus.
Refs.\cite{csikor,sasaki,takahashi}  claim the  existence  of a  bound
negative-parity  5Q  state, whereas  Ref.\cite{chiu}  claims that  the
lowest-lying 5Q bound state has positive-parity.
(Note that, below this positive-parity state, they actually obtained a
negative-parity state, which they claim to be a scattering state.)
Refs.\cite{mathur,ishii}  have  observed  no  evidence for  narrow  5Q
resonances  on  lattice.   Ref.\cite{okiharu}  studies the  static  5Q
potential to provide physical insights into effective models.
Under these circumstances,  our aim is to provide  an accurate lattice
QCD  result  using  anisotropic  lattice technique.   Furthermore,  we
propose a  new method with ``{\em hybrid  boundary condition (HBC)}'',
which  can  raise the  s-wave  $NK$  threshold  artificially.  HBC  is
expected  to   serve  as   a  convenient  tool   in  the   studies  of
negative-parity 5Q states.

\section{Lattice QCD Calculation}
\subsection{Results with the standard periodic BC}
We employ the standard  Wilson gauge action at $\beta=5.75$ ($a_s^{-1}
\simeq  1.1$ GeV)  on $12^3\times  96$ lattice  with  the renormalized
anisotropy   $a_s/a_t    =   4$   to   generate    504   gauge   field
configurations. The anisotropic lattice technique is known to serve as
a      powerful     tool      for      high-precision     measurements
\cite{matsufuru,nemoto,glueball}.  We use $O(a)$-improved Wilson quark
(clover)  action  \cite{matsufuru,nemoto}   adopting  four  values  of
hopping parameter $\kappa=0.1210(0.0010)0.1240$, which cover the quark
mass region as $m_s-2m_s$.  By keeping $\kappa_s=0.1240$ fixed for $s$
quark,  we change $\kappa=0.1210-0.1240$  for $u$  and $d$  quarks for
chiral  extrapolation.   We adopt  the  spatially  extended source  to
enhance the low-lying spectrum.

Assuming that the  quantum number of the $\Theta^{+}$  is spin $J=1/2$
and isospin $I=0$, we employ a {\em non-$NK$ type interpolating field}
\cite{sugiyama}
\begin{equation}
  \psi\equiv\epsilon_{abc}  \epsilon_{def} \epsilon_{cfg}
  \left(u_a^T C\gamma_5 d_b\right)
  \left(u_d^T C  d_e\right)
  C\bar{s}_g^T,
\end{equation}
to  construct  two-point 5Q  correlators.   Here,  $a-g$ denote  color
indices, and $C\equiv \gamma_4\gamma_2$ denotes the charge conjugation
matrix.  The Dirac  bispinor field $\psi$ couples to  baryon states of
both  parities.   Since  $\psi$  transforms  as  $\psi(t,\vec  x)  \to
+\gamma_4   \psi(t,-\vec  x)$  under   the  spatial   reflection,  the
projection  matrix $P_{\pm}\equiv  (1\pm \gamma_4)/2$  can be  used to
project the  intermediate states into positive  and negative parities,
respectively,  in  the  ``forward-propagation''  region  $0\ll  t  \ll
N_t/2$.

In both  the parity channels, the effective-mass  plots have plateaus,
where  the  single-exponential  fit  analysis is  performed.   In  the
positive-parity channel,  we obtain $m_{\rm 5Q}\simeq  2.25$ GeV after
the chiral extrapolation,  which is too massive to  be identified with
the    experimentally    observed    $\Theta^{+}(1540)$.     In    the
negative-parity channel,  we obtain $m_{\rm 5Q}\simeq  1.75$ GeV after
the  chiral extrapolation,  which  is rather  close  to the  empirical
value.

\subsection{Results with HBC}
In   order   to   identify   this   negative-parity   state   as   the
$\Theta^+(1540)$, it is  necessary to confirm that it  is a compact 5Q
resonance  state  rather than  an  $NK$  scattering  state.  For  this
purpose, we propose a new method with ``{\em hybrid boundary condition
(HBC)}''.  In the standard  boundary condition, the spatially periodic
BC is  imposed on $u$, $d$ and  $s$ quark fields.  Hence,  $N$ and $K$
are subject to  the spatially periodic BC, and  s-wave $NK$ scattering
spectrum starts at  $E_{\rm th}\simeq m_{N} + m_{K}$  as a consequence
of the negligibly weak $NK$  interaction in comparison with $m_{N}$ or
$m_{K}$.
In HBC,  the spatial BC is  twisted in a  flavor-dependent manner. The
spatially  periodic BC  is imposed  on  $s$ quark  field, whereas  the
spatially anti-periodic BC is imposed  on $u$ and $d$ quark fields. In
this case, since $N(uud, udd)$ and $K(u\bar{s}, d\bar{s})$ contain odd
number  of $u$  and  $d$ quarks,  they  are subject  to the  spatially
anti-periodic BC.  Due  to the finiteness of the  spatial box, allowed
spatial momenta of $N$ and $K$ are quantized as $p_i = (2n_i + 1)\pi /
L$, where $n_i \in \ZZ$, and  $L$ denotes the spatial extension of the
box. Note  that the momentum is  quantized in a  different manner from
the  standard periodic  BC case,  and $N$  and $K$  have  the non-zero
minimum momentum as $|\vec p_{\rm min}| = \sqrt{3}\pi/L$. As a result,
$NK$ scattering spectrum in HBC starts at an artificially raised value
as  $E_{\rm   th}  \simeq  \sqrt{m_{N}^2  +  \vec   p_{\rm  min}^2}  +
\sqrt{m_{K}^2  +  \vec  p_{\rm   min}^2}$  depending  on  the  spatial
extension of the box.
On  the other  hand, since  the $\Theta^+(uudd\bar{s})$  contains even
number  of  $u$ and  $d$  quarks, the  $\Theta^+$  is  subject to  the
spatially periodic BC.  Hence,  it can have the zero-spatial momentum.
If the  spatial size  of the $\Theta^+$  is sufficiently  smaller than
$L$, it is  safely expected that this compact 5Q  state should be less
sensitive to the change of the spatial BC. In this case, its mass will
stay at the same location.

\begin{figure}[t]
\begin{center}
\includegraphics[angle=-90,width=0.56\textwidth]{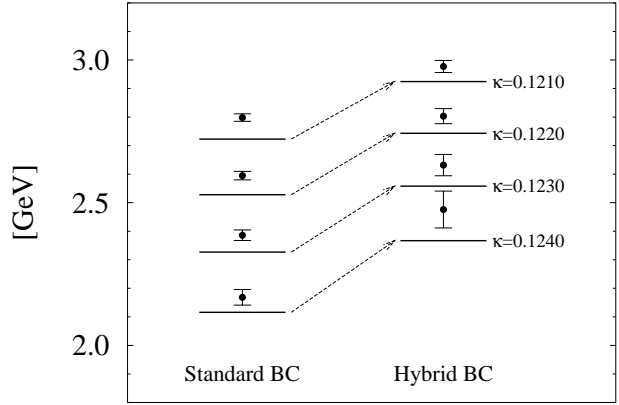}
\end{center}
\vspace{-3em}
\caption{Lattice QCD result for the negative-parity 5Q baryon.
The results  of standard periodic  BC (l.h.s.)  are compared  with HBC
(r.h.s.)  for each hopping parameter $\kappa$.  Circles denote results
obtained from  the single-exponential  fit analysis.  The  solid lines
denote the corresponding $NK$ thresholds $E_{\rm th}$.}
\label{hikaku}
\end{figure}
In \Fig{hikaku},  we show  the lattice QCD  result for 5Q  baryon.  We
compare the  HBC results with  the standard periodic BC  results.  The
closed  circles  denote  the  results of  the  single-exponential  fit
analysis.    The   solid   lines   denotes  the   corresponding   $NK$
thresholds. We see that the $NK$ thresholds are raised by 200--250 MeV
due to  HBC, and  that the  best fit masses  are raised  by consistent
amount as  the shifts of the  $NK$ thresholds.  This  implies that the
negative-parity  states with  the mass  $m_{\rm 5Q}  \simeq  1.75$ GeV
observed in the present lattice  QCD calculation is an $NK$ scattering
state rather than a compact 5Q resonance.

\section{Summary and Discussions}
To summarize, we have performed the anisotropic lattice QCD studies of
the 5Q  state for spin  $J^{P}=1/2^{\pm}$ and isospin $I=0$.   We have
obtained the lowest-lying positive-parity  state at $m_{\rm 5Q} \simeq
2.25$ GeV  after the chiral  extrapolation, which is considered  to be
too  massive   to  be  identified  with   the  $\Theta^+(1540)$.   The
lowest-lying  negative-parity  5Q  state  has been  found  at  $m_{\rm
5Q}\simeq 1.75$ GeV, which is  rather close to the empirical value. 
It  is  necessary  to clarify  whether  this  state  is a  compact  5Q
resonance state or an $NK$ scattering state in order to identify it as
the $\Theta^+(1540)$.
For this  purpose, we  have proposed a  new method with  ``{\em hybrid
boundary condition (HBC)}'', which can raise the s-wave $NK$ threshold
by a  few hundred MeV depending  on the extension of  the spatial box.
From  the HBC  analysis, it  has turned  out that  the negative-parity
state observed  at $m_{\rm  5Q} \simeq 1.75$  GeV is actually  an $NK$
scattering state.

In this  way, we  have observed  no clear signals  for the  compact 5Q
resonance   $\Theta^+(1540)$   both    in   the   negative   and   the
positive-parity channels.
Note  that another  null-result  is reported  by Ref.\cite{mathur}  in
lattice QCD.
Of  course, one  of the  possible implications  of these  two negative
results  is that  QCD may  not accommodate  the $\Theta^+(1540)$  as a
resonance   pole.    Note   that   experimental   existence   of   the
$\Theta^+(1540)$ has not yet been established so far.
As an interesting possibility, these null-results may be a consequence
of a  possibly complicated intrinsic  structures of the  $\Theta^+$ as
suggested by Refs.\cite{jaffe,karliner,bicudo,kishimoto}.
If this is the case,  the $\Theta^+(1540)$ may not be easily reachable
by the lattice QCD with a  simple 5Q interpolating field, and it would
be  desirable  to  introduce  series of  new  non-local  interpolating
fields, which can fit such non-trivial structures \cite{morningstar}.
Another  possible implication  is that  the $\Theta^+(1540)$  may have
simply       a      different      quantum       number.       Indeed,
Refs.\cite{hosaka,enyo,sarcar,nishikawa}  discuss  the possibility  of
spin $J=3/2$.   In this  respect, it would  be interesting  to perform
lattice QCD calculation for spin 3/2 states \cite{ishii-3/2}.
Of course, it is necessary to perform more systematic studies in order
to reveal the mysterious nature of the penta-quark $\Theta^+(1540)$.

\end{document}